\documentclass[useAMS,referee,12pt]{biom2}
\usepackage{amsmath, amsfonts, bm}
\usepackage{amssymb}
\usepackage{threeparttable}
\let\bibhang\relax
\usepackage{natbib}

\usepackage{graphicx,float}
\usepackage{multirow}

\usepackage{xcolor}

\def\bSig\mathbf{\Sigma}

\title[Interim Analysis of SMARTs for Survival Outcomes]{Interim Analysis in Sequential Multiple Assignment
Randomized Trials for Survival Outcomes}

\author{Zi Wang$^{1,*}$\email{ziw43@pitt.edu}, 
Yu Cheng$^{1,**}$\email{yucheng@pitt.edu}, and 
Abdus, S. Wahed$^{2,***}$\email{abdus\_wahed@urmc.rochester.edu} \\
$^{1}$Department of Statistics, University of Pittsburgh, Pittsburgh, Pennsylvania, U.S.A.\\
$^{2}$Department of Biostatistics and Computational Biology, University of Rochester, Rochester, New York, U.S.A.}

\begin{document}









\label{firstpage}


\begin{abstract}
Sequential multiple assignment randomized trials  mimic the actual treatment processes experienced by physicians and patients in clinical settings and inform the comparative effectiveness of dynamic treatment regimes. In such trials, patients go through multiple stages of treatment, and the treatment assignment is adapted over time based on individual patient characteristics such as disease status and treatment history. In this work, we develop and evaluate statistically valid interim monitoring approaches to allow for early termination of sequential multiple assignment randomized trials for efficacy targeting survival outcomes. We propose a weighted log-rank Chi-square statistic to account for overlapping treatment paths and quantify how the log-rank statistics at two different analysis points are correlated. Efficacy boundaries at multiple interim analyses can then be established using the Pocock, O'Brien Fleming, and Lan-Demets boundaries. We run extensive simulations to comparatively evaluate the operating characteristics (type I error and power) of our interim monitoring procedure based on the proposed statistic and another existing statistic. The methods are demonstrated via an analysis of a neuroblastoma dataset. 
\end{abstract}

%



\maketitle


%

\section{Introduction}
\label{s:intro}

Dynamic Treatment Regimes (DTRs), also known as adaptive treatment
strategies, are decision rules that recommend when and how to adjust treatment based on the treatment history \citep{Lavori2008, Chakraborty2014}. 
They are particularly effective in managing chronic diseases such as cancer, depression, and HIV, where multiple phases of treatment is common \citep{Murphy2005}. However, the vast array of potential DTRs can make it difficult to pinpoint the most effective regime.

Sequential Multiple Assignment Randomized Trials (SMARTs) have been proposed to mimic treatment decisions experienced by physicians and patients in clinical settings and provide a systematic approach to constructing and evaluating DTRs. In a SMART, each patient is randomly assigned to an initial treatment, and subsequent treatments at later stages are determined by patient's characteristics and intermediate outcomes observed in earlier stages \citep{Nahum-Shani2017}. Each stage in SMART corresponds to one of the critical decision points involved in the DTRs to be evaluated \citep{Bigirumurame2022}. An example is a 2-stage Neuroblastoma trial \citep{Matthay1999}, where patients were assigned to one of two initial treatments, and responders to the initial were further assigned to one of two maintenance therapies.

Although SMART is efficient for evaluating embedded DTRs, it typically takes longer to complete than traditional single-stage randomized control trials (RCTs) because of its sequential nature. The duration of SMART depends on the number of stages and the definition of intermediate and final outcomes studied in the design \citep{wu2021}. One way to improve the efficiency of SMART is to introduce interim monitoring (IM), which is often used in RCTs \citep{Jennsion1990, Ellenberg2002} to enable early trial termination if there is compelling evidence of treatment efficacy (superiority) or lack of efficacy (futility) \citep{Freidlin2010}.

IM procedures have been adopted for time-to-event outcomes for single-stage randomized control trials \citep{Kim1990, Tsiatis1982}. \cite{Tsiatis1982} developed the asymptotic joint distribution of test statistics from multiple interim analyses while \cite{Kim1990} developed IM procedure allowing for unequal increments between repeated analyses. \cite{Gu1999} performed interim analyses using Monte Carlo simulations that considered noncompliance, loss to follow-up, patient accrual patterns, and non-proportional hazard alternatives in clinical trial designs. \cite{Shen2003} proposed using a weighted average of standardized linear rank statistics for inference, along with a stopping rule that allows early termination if the experimental treatment shows no advantage or is inferior to the control. \cite{Chen2003} considered both survival and disease-free survival endpoints in their IM procedure: the trial would stop early if mortality showed a significant effect at the interim,  while the composite endpoint would be tested in the final analysis if the trial continued. Additionally, \cite{Broglio2014} developed methods to predict final trial results for survival outcomes based on interim data. All these approaches conduct interim analyses when a predetermined proportion of the total expected events have been observed.

Despite the wide use of interim analyses in single-stage trials, 
interim monitoring in SMARTs has been less studied due to the design's complexity.  Recent work has focused on continuous outcomes, with \cite{wu2021} proposing interim monitoring in SMART (IM-SMART), and \cite{Manschot2023} improving upon it to accommodate partial information. However, the interim monitoring method for time-to-event outcomes in SMART remains unclear. 
To address this gap, we develop various interim monitoring methods for SMARTs with survival outcomes. We adapt the log-rank test statistics proposed in \cite{Harrington1982} to compare survival functions among different DTRs. The development is nontrivial. First, log-rank statistics need to be carefully weighted to compensate for the systematic missingness due to sequential randomization. Second, we need to account for the correlation between log-rank statistics at different interim monitoring time points. Lastly, estimating the variance-covariance matrix of the log-rank test statistics is complex and requires special handling.  We also adopt an alternative test statistic recently proposed by \cite{Tsiatis2024}, referred to as Tsiatis-Davidian statistic, or in short, TD statistic in tables  and notation, and demonstrate that Tsiatis- Davidian statistic and the proposed weighted log-rank statistic belong to the same class of statistics. 

We consider various boundaries to compare the log-rank statistics at interim time points. \cite{Pocock1977} proposed constant boundaries across all analyses. However, interim monitoring using the Pocock boundaries tends to stop earlier and results in trials being stopped prematurely.
Thus, \cite{Brien1979} suggested using decreasing boundaries, where for a trial with $M$ analyses, the efficacy boundary at the $m$th interim analysis ($b_m$) is proportional to the boundary at the final analysis ($b_M$); More specifically,  $b_m/b_M=\sqrt{M/m}$.
One challenge in implementing these boundaries is that the covariance matrices of the test statistics depend on time, unlike the case with continuous outcomes as in \cite{wu2021}. We propose two ways to address this challenge. One is to use interim data to approximate covariance matrices at later time points and assume working independent increments. The other approach is to adopt the error spending functions proposed in \cite{Lan1983} that offer greater flexibility, as they do not require the total number or exact timing of interim analyses to be pre-specified. 
In this work, for SMARTs involving time-to-event outcomes, We provide essential tools to conduct interim monitoring using the above-mentioned boundaries.

The rest of this paper is organized as follows. Section \ref{sec2} describes the setting and notation for SMART design. In Section \ref{sec3}, we propose alternative test statistics and introduce interim monitoring procedures for SMART survival data under two SMART designs. Section \ref{sec4} shows the simulation results, and Section \ref{sec5} presents the analysis of  Neuroblastoma study data. Finally, we provide a discussion in Section \ref{sec6}.

\section{SMART Setting and Notation}
\label{sec2}
\subsection{Setting}
\begin{figure}
    \centerline{\includegraphics[scale=0.4]{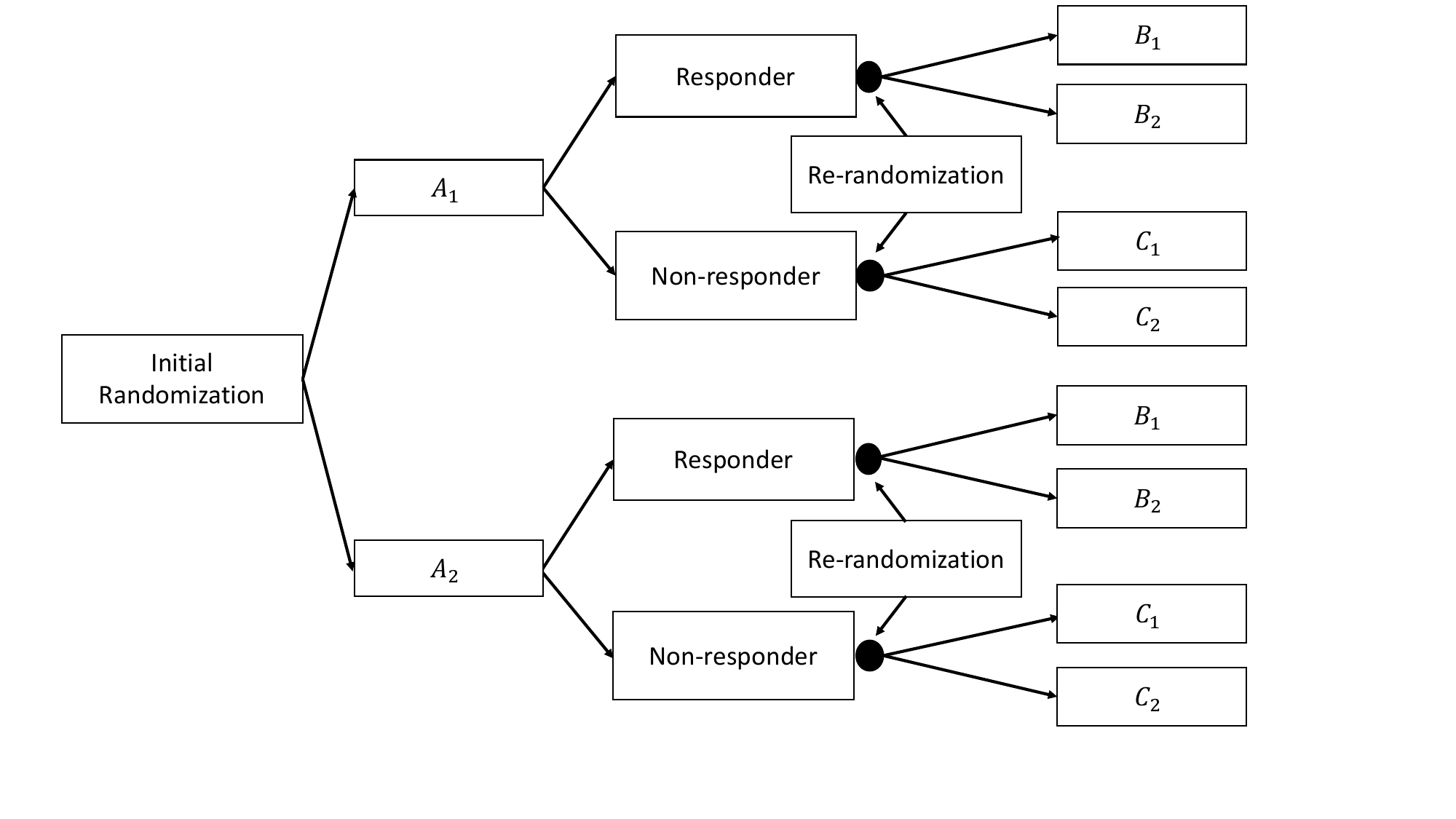}}
    \vspace{-.1in}
    \caption{A SMART Design with 8 Treatment Strategies.  \label{SMART_8Trt_BW}}
\end{figure}

In this project, for the development of the theoretical framework,  we first consider a two-stage SMART design (referred hereto forth as  SMART1) shown in Figure \ref{SMART_8Trt_BW}, where patients are initially randomized to treatment $A_1$ or $A_2$, and if they respond to the initial treatment, they are randomized to maintenance treatment $B_1$ or $B_2$ at the second stage; otherwise, they are randomized to salvage treatment $C_1$ or $C_2$. This SMART design enables the comparisons of eight embedded DTRs $A_jB_kC_l$, $j, k, l = 1,2$, where $A_jB_kC_l$ is a decision rule prescribing to treat the patient with the initial treatment $A_j$, and then by the maintenance therapy $B_k$ if the initial treatment was effective, and by the salvage treatment $C_l$, otherwise. We will consider a modified setting later, where non-responder (or responder) to the initial treatment do not receive further treatment (Section 3.5). The goal is to compare all embedded DTRs based on the distribution of the survival time, defined as the time from initial randomization to death.

\subsection{Observed Data}

For patient $i$, the generic treatment assignment indicator $I_i(x)$ equals $1$ if they are assigned to treatment $x$, and $0$ otherwise, and 
$\eta_i$ is the indicator of entering the second stage of the study. $T_{1i}$ is the time that patient $i$ spent in the first stage. $R_i$ is the response to the initial treatment. 
The observed time is $U_i=\min(T_i,V_i)$ with the event indicator  $\delta_i=I(T_i \leq V_i)$, where $T_i$ is the event time of interest, and $V_i$ is the censoring time. Thus, 
the observed data for SMART1 can be represented as  $$\{I_i(A_j), \eta_i, \eta_iT_{1i}, \eta_iR_i,\eta_iR_iI_i(B_k), \eta_i(1-R_i)I_i(C_l), U_i, \delta_i, j, k, l = 1, 2; \;i=1,\ldots,n\}, $$
where variables preceded by a binary indicator are observed only when that indicator assumes value $1$. For example, second stage treatment allocation $I_i(B_k)$ is observed if the patient moves to stage 2 ($\eta_i=1$) and responds ($R_i=1$).
\subsection{Inverse Probability Weights}
Log-rank test statistics are commonly used to compare the event time distribution across different groups. 
Had patients been randomly assigned to follow one of the eight DTRs, $A_jB_kC_l$, with $j,k,l = 1,2$, the analysis would have been straightforward through the application of log-rank test for independent groups. However, Such a trial is rare, as it requires a large sample size.
SMART is a more efficient design where DTRs share common paths, such as the same treatment path for responders but different paths for nonresponders. To account for this dependence and missingness due to subsequent randomizations, we need to weight each patient separately using inverse probability weighting  to properly utilise in the log-rank test statistics to compare embedded DTRs (Section 3). At an interim analysis, some newly enrolled participants may not have reached the second stage. To include them, we use a time-varying weight  with a nuanced definition of response status. Let $D_i(s)=\eta_iI(s \geq T_{1i})$ be the indicator of whether the response status is observed at time $s$. If the response status is not observed at time $s$ ($D_i(s)=0$), patients  assigned to $A_j$ receive a weight of $1/\ell_j$, where $\ell_j$ is the randomization probability for $A_j$. Once the response status is determined ($D_i(s)=1$),  if patient $i$ is a responder and assigned to $B_k$, the weight is further adjusted by $1/p_k$, and nonresponders assigned to $C_l$ have their weights multiplied by $1/q_l$  to reflect the second-stage randomization, where $p_k$ or $q_l$ are the randomization probabilities for $B_k$ or $C_l$, respectively.  
Therefore, following \cite{Guo2005}, we have the following time-dependent weight
\begin{equation}
W_{jkl,i}(s)= \frac{I_i(A_j)}{\ell_j}\left[1-D_i(s)+D_i(s)\left\{\frac{R_iI_i(B_k)}{p_k}+\frac{(1-R_i)I_i(C_l)}{q_l}\right\}\right]\label{eq:TDweightFor8}
\end{equation}
for a patient following the DTR $A_jB_kC_l, j,k,l=1,2.$
\subsection{Observed and Weighted Event and At-risk Processes}

Additional notation is needed before we introduce the weighted log-rank statistics. 
For patient $i$, the observed event process $N_i(s)=I(U_i \leq s, \delta_i=1)$ indicates an event at or before time $s$, and $N(s)=\sum_{i=1}^{n}N_i(s)$ is the total number of observed events by time $s$. If patient $i$ follows a specific DTR $A_jB_kC_l$, their data will be included in the computation of the relevant log-rank statistic with the  weight given in (\ref{eq:TDweightFor8}) to create a pseudo population where every patient follows this specific DTR.  Thus, the weighted number of events $ \bar{N}_{jkl}(s)=\sum_{i=1}^{n}W_{jkl,i}(s)N_i(s)$ represents the number of events at or before time $s$ if all patients were to follow the treatment strategy $A_jB_kC_l$. 
Similarly, let 
$Y_i(s)=I(U_i \geq s)$ be the at-risk process for patient $i$ and $Y(s)=\sum_{i=1}^{n}Y_i(s)$ be the total number of patients at risk at time $s$. $\bar{Y}_{jkl}(s)=\sum_{i=1}^{n}W_{jkl,i}(s)Y_i(s)$ is the weighted number of patients at risk by time  $s$ following the DTR $A_jB_kC_l$.

\section{Interim Monitoring in SMART}
\label{sec3}
\subsection{Weighted Log-rank Test Statistics}
Let $\Lambda_{jkl}(t)$ represent the cumulative hazard for the survival time of patients following the DTR $A_jB_kC_l$. Under the above setting, we would like to test the overall null hypothesis that there is no treatment effect. That is, $H_0$: the cumulative hazards for all DTRs are equal at any observable time, against the alternative hypothesis $H_1$: at least one cumulative hazard differs from others at some time point. 
Taking the DTR $A_1B_1C_1$ as the reference and comparing the cumulative hazards of other DTRs to the reference level, the null hypothesis can be written as $H_0:\gamma(t)=0$, where $\gamma(t)=\{\Lambda_{112}(t)-\Lambda_{111}(t), \Lambda_{121}(t)-\Lambda_{111}(t), \Lambda_{122}(t)-\Lambda_{111}(t),$
$\Lambda_{211}(t)-\Lambda_{111}(t), \Lambda_{212}(t)-\Lambda_{111}(t), \Lambda_{221}(t)-\Lambda_{111}(t), \Lambda_{222}(t)-\Lambda_{111}(t) \}^\intercal$. 
 We propose a Wald-type test statistic $T=n^{-1} Z^\intercal \widehat{\Sigma}^{-1} Z.$ Here, $Z=(Z_{jkl})^\intercal$ is a 7$\times$1 vector of weighted log-rank (LR) statistics, where 
 \begin{equation}
Z_{jkl}=\int_0^\infty \frac{\bar{Y}_{j k l}(s) \bar{Y}_{111}(s)}{\bar{Y}_{j k l}(s)+\bar{Y}_{111}(s)}\left\{\frac{\mathrm{d} \bar{N}_{j k l}(s)}{\bar{Y}_{j k l}(s)}-\frac{\mathrm{d} \bar{N}_{111}(s)}{\bar{Y}_{111}(s)}\right\},   \label{eq:LRfor8dtr}  
 \end{equation}
and $\widehat{\Sigma}$ is a consistent estimator of $\Sigma$, the asymptotic covariance matrix of $n^{-1/2}Z$.  
Let $Z_{jkl}(t)$ be $Z_{jkl}$ evaluated at time $t$, using only the information accrued up to time $t$, and $Z(t)=(Z_{jkl}(t))^\intercal$. Then, the test statistic evaluated at time $t$ is 
\begin{equation}
    T(t)=n^{-1} Z(t)^\intercal \widehat{\Sigma}^{-1}(t) Z(t),
    \label{eq:LRteststat}
\end{equation} 
where $\widehat{\Sigma}(t)$ is $\widehat{\Sigma}$ evaluated at time $t$. By the asymptotic normality of the weighted log-rank test statistic, when the null hypothesis is true,  $n^{-1/2}Z(t)$ converges to a multivariate normal distribution with mean 0 and covariance matrix $\Sigma(t)$. Therefore, by multivariate Slutsky's theorem, under $H_0$, $T(t) \sim \chi_{\nu}^{2}$ with degrees of freedom $\nu=rank(\Sigma_0(t))$, where $\Sigma_0(t)$ is $\Sigma(t)$ evaluated under $H_0$.

\subsection{Interim Monitoring Procedure}
Now, we are ready to discuss interim monitoring in SMART. Let $M$ be the total number of analyses, including the final analysis. We use $t_m$ to represent the decision time of the $m$th analysis and $n_m$ to denote the cumulative sample size up to time $t_m$. The test statistic in (\ref{eq:LRteststat}) at the $m$th analysis is denoted as $T(t_m)=n_m^{-1} Z(t_m)^\intercal \widehat{\Sigma}^{-1}(t_m) Z(t_m).$ 
Efficacy boundaries $b_1,\dots,b_M$ are defined such that the trial stops at the $m^{\star}$-th interim analysis  if the global null hypothesis is not rejected at all prior interim looks $1,\dots,m^{\star}-1$ and is rejected at the $m^{\star}$th look ($T(t_{m^{\star}})>b_{m^{\star}}$). The joint distribution of $T=(T(t_1),\dots, T(t_M))^\intercal$ is required to determine efficacy boundaries that maintain the overall type I error rate. That is,
$$
\sum_{m^*=1}^M \operatorname{pr}\left(\bigcap_{m=1}^{m^*-1} T\left(t_m\right) \leq b_m \cap T\left(t_{m^*}\right)>b_{m^*} \mid H_0\right)=\alpha.
$$
We specify the joint distribution of $T$  using the decomposition described in the following theorem:
\begin{theorem}
Let $Q(t_m)$ be a vector such that $T(t_m)=Q(t_m)^\intercal Q(t_m)$ for $m=1,\dots,M$. Under the null hypothesis, the stacked vector $Q=(Q(t_1),\dots,Q(t_M))^\intercal$ follows a multivariate normal distribution $Q \sim MN_{\nu \times M}(0,\Psi)$, where the matrix  $$\Psi=\left(\begin{array}{ccc}\Psi_{1,1} & \cdots & \Psi_{1, M} \\ \vdots & \ddots & \vdots \\ \Psi_{M, 1} & \cdots & \Psi_{M, M}\end{array}\right)$$ with $\Psi_{m,m^{\prime}}=\operatorname{cov}(Q(t_m),Q(t_{m^{\prime}}))$ and $\Psi_{m,m}=I_\nu$.
\end{theorem}

The proof easily follows from the results in \cite{dickhaus2015}. More specifically, note that we can write $T(t)=Z\left(t\right)^\intercal L\left(t\right) L\left(t\right)^\intercal Z\left(t\right)$, where $L(t)$ is the square root of $G(t)=n_{t}^{-1}(\Sigma(t))^g$. For $m=1,\dots,M$ and for any  $Q(t_m)$ satisfying $T(t_m)=Q(t_m)^\intercal Q(t_m)$, there exists an orthogonal matrix $P(t_m)$ such that $Q(t_m)=P(t_m)L(t_m)^TZ(t_m)$. Therefore, under $H_0$, $Q(t_m)$ has mean 0 and covariance matrix $P(t_m)P(t_m)^T=I_v$. Then following Definition 5.1 of \cite{dickhaus2015}, $T=(T(t_1),\dots,T(t_M))^\intercal$ follows a multivariate $\chi^2$ distribution of the Wishart type with an associated correlation matrix $\Psi$.

\subsection{Calculation of the Correlation Matrix $\Psi$}
Theorem 1 only specifies the diagonal blocks of the matrix $\Psi$. To calculate the whole matrix $\Psi$, we apply singular value decomposition to $G(t)=n_{t}^{-1}(\Sigma(t))^g$, i.e., $G(t)=U(t)Q(t)U(t)^\intercal$. Let $L(t)=U(t)Q(t)^{\frac{1}{2}}$ and $Q\left(t\right)=L\left(t\right)^{\intercal} Z(t)$, where $Z(t)$ is defined in Section 3.1. Then, $G(t)=L\left(t\right) L\left(t\right)^\intercal$ and 
$
T\left(t\right)=n_t^{-1} Z\left(t\right)^\intercal\left(\Sigma\left(t\right)\right)^g Z\left(t\right)= 
 Z\left(t\right)^\intercal L\left(t\right) L\left(t\right)^\intercal Z\left(t\right)=Q\left(t\right)^\intercal Q\left(t\right).
$ Thus, $T(t_m)=Q(t_m)^\intercal Q(t_m)$ for $m=1,\dots,M$, and when $t_m<t_{m^{\prime}}$, $\Psi_{m,m^{\prime}}=
\operatorname{cov}(Q(t_m),Q(t_{m^{\prime}}))=L(t_m)^{\intercal}\operatorname{cov}(Z(t_m),Z(t_{m^{\prime}}))L(t_{m^{\prime}})$.
Therefore, it suffices to calculate $\widehat{\operatorname{cov}}(Z(t))=n\widehat{\Sigma}(t)$ and $\widehat{\operatorname{cov}}(Z(t_m),Z(t_{m^{\prime}}))$. We propose to estimate these two quantities by asymptotic linearization.


Since $d\bar{N}_{j k l}(s)=\sum_{i=1}W_{jkl,i}(s)dN_i(s)$, the log-rank statistics in equation (\ref{eq:LRfor8dtr}) can be expressed as
$$Z_{jkl}=\sum_{i=1}^n\int_0^\infty  \frac{\bar{Y}_{111}(s)W_{jkl,i}(s)-\bar{Y}_{jkl}(s)W_{111,i}(s)}{\bar{Y}_{j k l}(s)+\bar{Y}_{111}(s)}dN_i(s).$$
Recall $\bar{Y}_{jkl}(s)$ is defined as $\bar{Y}_{jkl}(s)=\sum_{i=1}^nW_{jkl,i}(s)Y_i(s)$, we have
$\sum_{i=1}^n\{\bar{Y}_{111}(s)W_{jkl,i}(s)-\bar{Y}_{jkl}(s)W_{111,i}(s)\}/\{\bar{Y}_{j k l}(s)+\bar{Y}_{111}(s)\}Y_i(s)=0.$
Let $dM_i(s)=dN_i(s)-Y_i(s)d\Lambda_0(s)$ denote the residual for patient $i$. Then,
$$Z_{jkl}=\sum_{i=1}^n\int_0^\infty  \frac{\bar{Y}_{111}(s)W_{jkl,i}(s)-\bar{Y}_{jkl}(s)W_{111,i}(s)}{\bar{Y}_{j k l}(s)+\bar{Y}_{111}(s)}dM_i(s), $$
where $d\Lambda_0(s)$  is the overall hazards when $H_0$ is true. Under regularity conditions, $\bar{Y}_{jkl}(s)/\{\bar{Y}_{j k l}(s)+\bar{Y}_{111}(s)\}$ converges to a deterministic process, which is denoted as $\pi_{jkl}(s)$. Then, $Z_{jkl}$ is asymptotically equivalent to the sum of $n$ independent terms 
\begin{equation*}
\begin{split}
\frac{1}{\sqrt{n}}Z_{jkl} &= \frac{1}{\sqrt{n}}\sum_{i=1}^n \int_0^\infty \{(1-\pi_{jkl}(s))W_{jkl,i}(s)-\pi_{jkl}(s)W_{111,i}(s)\}dM_i(s)+o_p(1) \\ 
&= \frac{1}{\sqrt{n}}\sum_{i=1}^nZ_{jkl,i}+o_p(1).
\end{split}
\end{equation*}

where $$Z_{jkl,i}=\int_0^\infty \{(1-\pi_{jkl}(s))W_{jkl,i}(s)-\pi_{jkl}(s)W_{111,i}(s)\}dM_i(s).$$


It is easy to show that $Z_{jkl,i}$ has mean 0 under $H_0$ \citep{Tsiatis2024}. Consider the vector associated with patient $i$, $\hat{Z}_i=(\hat{Z}_{jkl,i})^\intercal \in \mathbb{R}^7 $, where  $$\hat{Z}_{jkl,i}=\int_0^\infty \frac{\bar{Y}_{111}(s)W_{jkl,i}(s)-\bar{Y}_{jkl}(s)W_{111,i}(s)}{\bar{Y}_{j k l}(s)+\bar{Y}_{111}(s)}\left\{dN_i(s)-Y_i(s)\frac{dN(s)}{Y(s)}\right\}$$ is an estimator of $Z_{jkl,i}$.
Here, $\int_0^t dN(s)/Y(s)$ is the Nelson-Aalen estimator of $\Lambda_0(t)$. Let $\hat{Z}_i(t)=(\hat{Z}_{jkl,i}(t))^\intercal$ be $\hat{Z}_i$ evaluated at time $t$. Then, $\Sigma(t)$, which is the asymptotic covariance matrix of $n^{-1/2}Z(t)$, can be estimated by $n^{-1}\sum_{i=1}^n\hat{Z}_i(t)\hat{Z}_i(t)^\intercal.$

Regarding the estimation of  $\operatorname{cov}(Z(t_m),Z(t_{m^{\prime}}))$ - had we known the full data at the time of interim analysis, since  $\operatorname{cov}(Z_{jkl,i}(t_m),Z_{jkl,i^{\prime}}(t_{m^{\prime}}))=0 \textrm{\ for } i \neq i^{\prime},$ and 
$\operatorname{cov}(Z_{jkl,i}(t_m),Z_{jkl,i}(t_{m^{\prime}})) =E(Z_{jkl,i}(t_m)Z_{jkl,i}(t_{m^{\prime}}))$ could be estimated by $Z_{jkl,i}(t_m)Z_{jkl,i}(t_{m^{\prime}})$, an estimator of $\operatorname{cov}(n_m^{-1/2}Z(t_m),$ \\$n_{m^{\prime}}^{-1/2}Z(t_{m^{\prime}}))$ when $t_m < t_{m^{\prime}}$ would have been obtained as
$(n_mn_{m^{\prime}})^{-1/2}\sum_{i=1}^{n_m}\hat{Z}_{i}(t_m)\hat{Z}_{i}(t_{m^{\prime}})^\intercal.$

However, in practice, the full data is not available at the interim analysis time. In order to determine the efficacy boundaries at the interim analyses without the full data, we consider two approaches and illustrate them with $M=2$. One approach is to assume the independent increment property and $\operatorname{cov}(Z(t_1))/n_1 \approx \operatorname{cov}(Z(t_2))/n_2$ when we choose $t_1$ such that we expect to observe half of the events by time $t_1$. 
The second approach is to specify an error spending function $\alpha^{\star}(t)$ and determine the boundaries sequentially \citep{Lan1983}. Here, $\alpha^{\star}(t)$ is a strictly increasing function; boundaries $b_1$ and $b_2$ satisfy $\operatorname{pr}(T(t_1)>b_1)=\alpha^{\star}(t_1)$ and $\operatorname{pr}(T(t_1)\leq b_1,T(t_2)>b_2)=\alpha - \alpha^{\star}(t_1)$. According to \cite{Lan1983}, we may choose $\alpha_1^{\star}(t)=2(1-\Phi(z_{\alpha/2}/(t/L)^{1/2})$ and $\alpha_2^{\star}(t)=\alpha\operatorname{log}(1+(e-1)t/L)$ 
to generate boundaries that are similar to the Pocock and OBF type boundaries with $L=2t_1$, where $\Phi$ is the cumulative distribution function of a standard normal random variable. We will refer to these boundaries as LD-Pocock and LD-OBF respectively.

\subsection{Alternative test statistics}
Recently, \cite{Tsiatis2024} proposed a different test statistic based on the score test for a Cox proportional hazards model. Their statistic, abbreviated as the TD statistic, is based on $Z^{TD}=(Z_{jkl}^{TD})^\intercal$, where \begin{equation*}
Z_{jkl}^{TD}=\sum_{i=1}^n\int_0^\infty W_{jkl,i}(s)\left\{dN_i(s)-Y_i(s)d\hat{\Lambda}_0^{TD}(s)\right\},    
\label{TDzjkl}
\end{equation*}
with $d\hat{\Lambda}_0^{TD}(s)= \sum_{j,k,l =1,2}d\bar{N}_{jkl}(s)/\sum_{j,k,l =1,2}\bar{Y}_{jkl}(s)$ being the estimated overall hazards under $H_0$. This  statistic is equivalent to the statistic used in \cite{Li2014} under the SMART setting.

The weighted log-rank statistic defined in  (\ref{eq:LRfor8dtr}) can be written as \begin{equation*}Z_{jkl}=\sum_{i=1}^n\int_0^\infty W_{jkl,i}(s)\left\{dN_i(s)-Y_i(s)d\hat{\Lambda}_0(s)\right\},\label{altLR}\end{equation*}
where $d\hat{\Lambda}_0(s)=\{d\bar{N}_{111}(s)+d\bar{N}_{jkl}(s)\}/\{\bar{Y}_{111}(s)+\bar{Y}_{jkl}(s)\}.$
This implies that our statistic and Tsiatis-Davidian statistic 
belong to the same class except that they use different estimators for the common hazard function under the null hypothesis of no difference. While \cite{Tsiatis2024} averages the event and risk processes across all DTRs, the LR statistic defined above averages the same over the DTR involved ($A_jB_kC_l$) and a reference DTR (more specifically, $A_1B_1C_1$). 

\subsection{Modified SMART}

In some SMARTs, patients are initially randomized to receive treatment $A_1$ or $A_2$. Responders to the initial treatment receive
maintenance treatment $B_1$ or $B_2$ at the second stage, while non-responders are not further randomized. We refer to this design as SMART2. The goal is to compare four embedded DTRs $A_jB_k$, for $j,k=1,2$, where $A_jB_k$ stands for ``Treat with $A_j$ initially, followed by $B_k$ if they respond." 

In this case, the observed data become $\{I_i(A_j), \eta_i, \eta_iR_i,\eta_iR_iI_i(B_k),U_i, \delta_i\}, i=1,\ldots,n$. The observed event process is defined as $N_i(s)=I\left(U_i \leq s, \delta_i=1\right)$, and the observed at-risk process becomes $Y_i(s)=I(U_i\geq s)$.
The time-dependent weight for treatment strategy $A_jB_k$ is defined as
$W_{j k, i}(s)=I_i\left(A_j\right)/\ell_j\left[1-D_i(s)+D_i(s)\left\{R_i I_i\left(B_k\right)/p_k+\left(1-R_i\right)\right\}\right].$ 
The corresponding weighted number of events by time $s$ for $A_jB_k$ is $N_{j k}(s)=\sum_{i=1}^n W_{j k, i}(s) N_i(s)$.  Similarly, we can define the weighted at-risk process  $Y_{jk}(s)$.

Let $\Lambda_{jk}(t)$ be the cumulative hazard with respect to treatment strategy $A_jB_k$. The null hypothesis can be written as $H_0:(\Lambda_{12}(t) - \Lambda_{11}(t),\Lambda_{21}(t) - \Lambda_{11}(t), \Lambda_{22}(t) - \Lambda_{11}(t))^{\intercal}=0,$ for any $t$. 
The weighted log-rank statistic for testing $H_0$ is $Z=(Z_{12},Z_{21},Z_{22})^{\intercal}$, where
$
Z_{jk}=\int_0^
\infty \bar{Y}_{j k}(s) \bar{Y}_{11}(s)/\{\bar{Y}_{j k}(s)+\bar{Y}_{11}(s)\}\{d \bar{N}_{j k}(s)/\bar{Y}_{j k}(s)-d \bar{N}_{11}(s)/\bar{Y}_{11}(s)\}.
$
The Wald type test statistic can be similarly defined as in equation  (\ref{eq:LRteststat}). Under $H_0$, $T \sim \chi_{\nu}^2$ with df $\nu=3$.

\section{Simulation}
\label{sec4}
\subsection{Data generation}
We evaluate the performance of the proposed interim analyses for two SMART designs,  SMART1 design outlined in Section 2.1 involving eight DTRs,  and  SMART2 design outlined in Section 3.5 involving four DTRs. For  SMART1, we generated $1000$ datasets, each with $n=500$ patients arriving independently and uniformly over $5$ years. Initial treatment indicator was generated from a Bernoulli distribution with equal probability ($Ber(0.5)$). We also generated the second stage potential advancement indicator $\eta_i \sim Ber(p_\eta)$.
When $\eta_i=0$ (death in stage 1), time in stage 1 $T_{1i}$ was generated from $Exp(\theta_j^N)$, and whereas when $\eta_i$ = 1, we generated a response indicator 
$R_i [Ber(p_R)]$, and $T_{1i} [Exp(\theta_j)]$ with $p_R=0.6$. For responders ($R_i = 1$), maintenance treatment was assigned through  $Ber(0.5)$. Time in stage 2, $T_{2i}$, was generated from $Exp(\theta_{jk}^{R})$. For non-responders ($R_i=0$), salvage treatment was assigned with equal probability, and $T_{2i}$ was generated from an $Exp(\theta_{jl}^{NR})$ distribution. The total survival time was then defined as  $T_i=(1-\eta_i)T_{1i}+\eta_i(T_{1i}+T_{2i})$. 
For the SMART2 design, the simulation was similar except that for non-responders there was no second stage treatment assignment, and their time in stage 2 was generated from $Exp(\theta_{j}^{NR})$.  Censoring time, $V_i$ $[Unif(0,\nu)]$, was generated independent of the survival times.\\ 

\subsubsection{Null scenarios} 
The following parameters were used for the SMART1 design : $\theta_j^N=(5,5), \theta_j = (5,5), \theta_{j k}^{R}=(5,5,5,5)$, and $\theta_{jl}^{N R}=(5,5,5,5)$. For the SMART2 design, we used $\theta_j^N=(3,3), \theta_j = (3,3), \theta_{j k}^{R}=(2,2,2,2)$, and $\theta_{jl}^{N R}=(5,5)$. The parameter $\nu$ was controled so that censoring proportion was approximately 10\%, 20\%, or 40\%. We also considered two values for $p_{\eta}$, namely, $0.90$ and $0.75$, the results for the latter shown in the supplementary materials. We conducted a single interim analysis when 50\% of the events were observed (i.e., the information proportion was 0.5).

\subsubsection{Alternative scenarios}
Next, we evaluated the performance under four alternative scenarios. The first two scenarios pertain to the SMART1 design: (1) $\theta_j^N=(5,5), \theta_j = (5,5), \theta_{j k}^{R}=(2, 4, 3, 4), \theta_{jl}^{N R}=(3.2,3,2.9,2),\nu = 2.5$; and (2) $\theta_j^N=(5,5)$, $\theta_j = (5,5), \theta_{j k}^{R}=(2.8, 4.6, 2.3,4.9), \theta_{jl}^{N R}=(5.8, 4.3, 5.2, 6.5),\nu = 2.1$. The last two are for the SMART2 design: (3) $\theta_j^N=(3,3)$, $\theta_j = (3,3), \theta_{j k}^{R}=(2,3.2,2.5,4), \theta_{jl}^{N R}=(6,6),\nu= 2.9$; and (4) $\theta_j^N=(3,3), \theta_j = (3,3), \theta_{j k}^{R}=(2.7,6,4.9,3)$, $\theta_{jl}^{N R}=(3.8,7.2), \nu = 2.8$. The parameters of SMART2 indicate that there is no further assignment for non-responders. The values of $\nu$ were selected for each scenario to control the censoring rates at approximately 20\%.

\subsection{Boundary calculation}
To implement the proposed IM procedures, we need to determine the efficacy boundaries based on the distribution of the quadratic form of a multivariate normal random vector $Q$, with mean zero and the covariance matrix $\Psi$ as shown in Theorem 1. Thus, the key component here is to estimate $\Psi$. Since in the simulations we know the true data generatiing mechanism, we approximate $\Psi$ by generating a large SMART dataset of 10,000 individuals based on the parameters specified in the above null scenarios. We then obtain $B=100,000$ replicates of $Q$ vectors sampled from a multivariate normal distribution with the estimated covariance matrix. The corresponding quantiles of the empirical joint distribution of $T(t_m)=Q(t_m)^{\intercal} Q(t_m)$ for $m=1,2$ are taken as the efficacy boundaries. For SMART studies in practice, $\Psi$ is estimated based on the observed data accumulated up to the interim analysis, and the covariance matrix in the final analysis can be approximated based on the interim one, as discussed in Section 3.3, resulting in either Pocock or OBF boundary. Alternatively, one can use the Lan and Demets (LD) procedure with the alpha spending function, where $\Psi$ is estimated based on the data accumulated at each decision point. The resulting  boundaries are labeled as LD-Pocock or LD-OBF boundaries. 
All these efficacy boundaries calculated using various approaches are presented in Table \ref{EB_Sim}. The first two rows display boundaries computed using full data at the interim stage, referred to as ``oracle" boundaries, while the third and fourth rows show boundaries derived by approximating the final covariance matrix using that at interim. The table reveals that, for the SMART2 design, the efficacy boundaries for the weighted log-rank and Tsiatis-Davidian statistics are similar. However, for the SMART1 design, the boundaries differ between the two,  the discrepancy arising from the difference in the degrees of freedom (for SMART1 design, the weighted log-rank statistic has seven degrees of freedom, while Tsiatis-Davidian statistic has five degrees of freedom). We also evaluated efficacy boundaries under various other null scenarios and found that the efficacy boundaries stayed the same regardless of the settings.

\begin{table}
\centering
\caption{Efficacy boundaries for 2 SMART Designs.Oracle: boundaries calculated using full data at interim. LD-Pocock and LD-OBF: Pocock and OBF boundaries obtained using a particular error spending function recommended by \cite{Lan1983}. LR and TD approaches are our proposed weighted log-rank statistic and the statistic proposed in \cite{Tsiatis2024}, respectively.}
\label{EB_Sim}
\begin{threeparttable}
\begin{tabular}{cccccc}
\hline
 & \multicolumn{2}{c}{ SMART1 } & & \multicolumn{2}{c}{SMART2} \\
 \cline{2-3} \cline{5-6}
  & LR & TD & & LR & TD \\
\hline
 Oracle  Pocock & (15.74, 15.74) & (12.46, 12.46) & & (9.03, 9.03) & (9.08, 9.08)  \\
  Oracle OBF & (20.11, 14.22) & (15.85, 11.21) &  & (11.43, 8.08) & (11.43, 8.08)  \\
   Pocock & (15.98, 15.98) & (12.79, 12.79) &  & (9.38, 9.38) & (9.42, 9.42)   \\
  OBF & (19.80, 14.00) & (15.63, 11.05) & & (11.10, 7.85) & (11.07, 7.83)  \\
  LD-Pocock & (15.43, 16.14) & (12.25, 12.70) & & (8.85, 9.24)  & (8.87, 9.31)    \\
  LD-OBF  & (20.03, 14.22) & (16.35, 11.16) & & (12.61, 7.95)  & (12.72, 7.95) \\
\hline
\end{tabular}

\end{threeparttable}
\end{table}

\subsection{Simulation Results}
To investigate the performance of our interim monitoring approach, we generated datasets with the same set of parameters above. 
We used two methods to estimate the covariance matrices of $Z$: the proposed method, which derives  $\operatorname{cov}(Z(t))$ and $\operatorname{cov}(Z(t_m),Z(t_{m^{\prime}}))$ using asymptotic linearization (Refer to Section 3.3), and a bootstrap estimator, which estimates $\operatorname{cov}(Z(t))$ and $\operatorname{cov}(Z(t_m),Z(t_{m^{\prime}}))$ from 100 bootstrapped samples. \\

\subsubsection{Simulation Results: Null Scenarios}
Table \ref{Trt_Null} shows the results under the null scenario for censoring rate 20\%, and 90\% of the patients entering the second stage.  ``Rej at interim" is the estimated probability of rejection at the interim analysis, and ``Rej at final" is the probability of rejection at the final analysis conditional on not being rejected at the interim analysis. We also estimated the overall rejection rate when the null hypothesis is true, denoted as ``Type I Error" in Table \ref{Trt_Null}. Performance under $n=1000$ was also investigated, but the results are not shown here due to space limitations. The first row in the table is for a single analysis (regular SMART) at the conclusion of the trial and the remaining rows are for the case when there are two total analysis (single interim analysis). We observe that the type I error rates of the LR and TD statistics using the proposed covariance estimation method are well-controlled near the nominal 0.05 level across all boundary types. However, the empirical type I error rates from the bootstrap procedure are mostly inflated. 
We observe similar patterns under 10\% censoring (Web Table 2) and 40\% censoring (Web Table 3) presented in the Supplementary material. Also presented in the supplementary materials are the results for the case when 75\% of patients enter the second stage (Web Tables 4, 5 and 6). Higher censoring rates tend to result in slightly lower type I errors, making the decision more conservative. Bootstrap estimators largely yield inflated type I errors  across scenarios. Moreover, TD estimators have slightly inflated type I errors than LR estimators for SMART1, especially under 10\% and 20\% censoring. Both estimators maintain type I errors well for SMART2. The discrepancy observed in SMART1 may be explained by the different degrees of freedom for the two statistics (7 df for LR and 5 df for TD), while both statistics have 3 df in SMART2.\\

\begin{table}
\centering
\caption{Simulation results for two SMART designs under Null Scenarios with censoring rate approximately 20\% and 90\% patients entering stage 2. $n = 500$.M: Number of total analysis including the final. Rej at interim: rejection rate at the interim analysis. Rej at final: the probability of rejection at the final analysis condition on not rejecting at the interim analysis. Oracle: boundaries calculated using full data at interim. LR and TD: the proposed weighted log-rank and \cite{Tsiatis2024}  statistics normalized using the proposed covariance estimates. BLR and BTD: the statistics normalized using the bootstrap covariance estimates. LD-Pocock and LD-OBF: Pocock and OBF boundaries obtained using a particular error spending function recommended by \cite{Lan1983}. Results are shown in percentages within 1 decimal place.} 
\label{Trt_Null}
\resizebox{0.9\textwidth}{!}{
\begin{threeparttable}
\begin{tabular}{@{}cccccccccrcr@{}}
\hline
\multirow{2}{*}{M} & \multirow{2}{*}{Boundary} & \multirow{2}{*}{Rejection Rate 
 } & \multicolumn{4}{c}{ SMART1 } & & \multicolumn{4}{c}{ SMART2 } \\
\cline{4-7} \cline{9-12}
&  &  & LR & BLR & TD & BTD & & LR & BLR & TD & BTD \\
\hline
1 & NA & Type I Error & 3.8 & 3.5 & 6.1 &  6.9 & &5.4 & 9.1 &4.8 & 8.6 \\
\hline
\multirow{18}{*}{2} & Oracle & Rej at interim & 2.3 & 3.2 & 4.3 & 5.8 & & 3.0 & 7.0 & 3.1 &  7.1\\ 
 & Pocock  & Rej at final & 1.8 & 1.7 & 2.8 & 3.1 & &3.0 & 4.3 & 2.9 & 4.5  \\  
 &  & Type I Error & 4.1 & 4.8 & 7.0 & 8.9 & & 5.9 & 11.1 & 6.0  & 11.3 \\
 & Oracle & Rej at interim & 0.3 & 0.6 & 1.6 & 1.7 & & 1.3 & 3.2 & 1.1 & 2.7 \\  
 & OBF & Rej at final  & 3.7 & 3.2 & 4.8 & 6.7 & & 4.2 & 6.9 & 4.1 & 7.2 \\
 &  &  Type I Error  & 4.0 & 3.8 & 6.4 & 8.3 & & 5.5 & 9.9 & 5.2 & 9.8 \\
 \cline{2-12}
 & Pocock & Rej at interim & 2.2 & 2.9 & 4.0 & 4.9 & & 2.8 & 5.9 & 2.7 & 6.4 \\  
 &  & Rej at final  & 1.5 & 1.6 & 2.7 & 2.7 & & 2.7 & 4.2 & 2.8 & 4.3 \\
 &  &  Type I Error  & 3.7 & 4.5 & 6.6 & 7.5 & & 5.4 & 9.9 & 5.5 & 10.5 \\
 & OBF & Rej at interim & 0.5 & 0.7 & 1.7 & 2.2 & & 1.4 & 3.6 & 1.4& 3.3 \\  
 &  & Rej at final  & 3.8 & 3.6 & 5.0 & 6.6 & & 4.9 & 7.7 & 4.5 & 7.6 \\
 &  &  Type I Error  & 4.3 & 4.3 & 6.6 & 8.7 & & 6.2 & 11.1 & 5.8 & 10.7 \\ 
 \cline{2-12}
  & LD-Pocock & Rej at interim  & 3.0 & 3.8 & 4.6 & 6.2 & &3.3 & 7.2 & 3.4 & 7.8 \\ 
   & & Rej at final  & 1.4 & 1.4 & 2.4 & 2.7 & & 2.8 & 4.2 & 2.7 & 4.1  \\
 &  &  Type I Error  & 4.4 & 5.1 & 6.9 & 8.7 &  & 6.0 & 11.1 & 6.0 & 11.6 \\ 
 & LD-OBF & Rej at interim & 0.3 & 0.6 & 1.4 & 1.5 & & 1.0 & 2.1 & 0.8 & 2.3 \\  
 &  & Rej at final  & 3.7 & 3.2 & 5.1 & 6.7 & & 4.7 & 7.6 & 4.3 & 7.5 \\
 &  &  Type I Error  & 4.0 & 3.8 & 6.4 & 8.1 & & 5.7 & 9.6 & 5.1 & 9.7 \\
 \hline
\end{tabular}

\end{threeparttable}}
\end{table}

\subsubsection{Simulation Results: Alternative Scenarios}
Due to the inflated type I error rates observed with the bootstrapped procedure under the null scenarios, we assessed performance of the proposed interim approach using only the proposed covariance estimates for the alternative scenarios. The results are summarized in Table \ref{Trt_Alt}, which also includes the expected sample size [$E(n)$] when a trial stops. We first observe that trials tend to stop earlier with the Pocock boundaries at the cost of lower power compared to a single analysis. The OBF and the LD-OBF maintain power (similar to a single analysis) across all scenarios, with a notable reduction in the average sample size . For instance,  under Alternatives 1-4, the expected sample size with OBF boundaries ranges from 406 to 467, 
representing a reduction of 7\% to 19\% compared to the planned total sample size of 500.  Furthermore, the TD methods achieve higher power than the LR statistic for SMART1, and the powers are similar for SMART2. This difference in power may be explained by the slightly inflated type I errors that we've observed for the TD estimator in SMART1.
Additional simulations presented in the supplementary materials with censoring rates 10\% (Web Table 7) and 40\% (Web Table 8), and 75\% of patients entering the second stage (Web Tables 9 and 10) also support similar conclusions. 

\begin{table}
\centering
\caption{Simulation results for SMART1 and SMART2 under alternative scenarios with a censoring rate of approximately 20\%, $n=500$.M: Number of total analysis including the final. Rej at interim: rejection rate at the interim analysis. Rej at final: the probability of rejection at the final analysis condition on not rejecting at the interim analysis. Oracle: boundaries calculated using full data at interim. LR and TD: the proposed weighted log-rank and \cite{Tsiatis2024}  statistics normalized using the proposed covariance estimates. LD-Pocock and LD-OBF: Pocock and OBF boundaries obtained using a particular error spending function recommended by \cite{Lan1983}. All results (except $E(n)$) are shown in percentages rounded to integers.} 
\label{Trt_Alt}
\resizebox{0.9\textwidth}{!}{
\begin{threeparttable}
\begin{tabular}{@{}cccccccccccccc@{}}
\hline
\multirow{3}{*}{M} & \multirow{3}{*}{Boundary} & \multirow{3}{*}{ Rejection Rate } & \multicolumn{5}{c}{ SMART1 } & & \multicolumn{5}{c}{ SMART2 } \\
\cline{4-8}  \cline{10-14}
 &  &  & \multicolumn{2}{c}{Alt1} & & \multicolumn{2}{c}{Alt2} & & \multicolumn{2}{c}{Alt3} & & \multicolumn{2}{c}{Alt4} \\
 \cline{4-5} \cline{7-8} \cline{10-11} \cline{13-14}
 &  &  & LR & TD & & LR & TD & & LR & TD & & LR & TD \\
\hline
1 & NA & Power & 78 & 82 & & 92 & 95 & & 70 & 70 & & 91 & 91 \\
\hline
\multirow{24}{*}{2} & Oracle & Rej at interim & 33 & 42 & & 47 & 60 & & 28 & 28 & & 50 & 49 \\ 
 &  & Rej at final & 60 & 64 & & 78 & 82 & & 53 & 51 & & 75 & 76  \\  
 &  & Power & 73 & 79 & & 88 & 93 & & 66 & 65 & & 88 & 87  \\
 &  & E(n) & 431 & 413 & & 402 & 375 & & 441 & 441 & & 395 &  398   \\ 
 & OBF & Rej at interim & 16 & 26 & & 26 & 43 & & 14 & 16 & & 31 & 32\\  
 &  & Rej at final & 73 & 76 & & 88  & 91 & & 65 & 63 & & 86 & 85  \\
 &  & Power & 78 & 82 & & 91 & 95 & & 70 & 69 & & 90 & 90   \\
 &  & E(n) & 466 & 446 & & 446 & 409 & & 471 & 468 & & 435 & 432 \\ 
\cline{2-14}
 & Pocock & Rej at interim & 32 & 40 & & 45 & 57 & &26 & 26 & & 47 & 46  \\  
 &  & Rej at final & 59 & 64 & & 78 & 81 & & 49 & 49 & & 73 & 73 \\
 &  & Power & 72 & 78 & & 88 & 92 & & 63 & 62 & & 86 & 85   \\
 &  & E(n) & 434 & 415 & & 406 & 380 & & 446 & 446 & & 402 & 404  \\ 
 & OBF & Rej at interim & 17 & 27 & & 27 & 45 & & 16 & 17 & & 33 & 35  \\  
 &  & Rej at final & 74 & 76 & & 89 & 91 & & 65 & 65 & & 86 & 86  \\
 &  & Power & 79 & 83 & & 92 & 95 & & 70 & 71 & & 91 & 91  \\
 &  & E(n) & 465 & 443 & & 443 & 406 & & 467 & 465 & & 431 & 427 \\ 
\cline{2-14}
 & LD-Pocock & Rej at interim & 34 & 43 & & 49 & 62 & & 29 & 29 & & 51 & 50 \\  
 &  & Rej at final & 57 & 63 & & 77 & 89 & & 50 & 49 & & 73 & 73  \\
 &  & Power & 72 & 79 & & 88 & 92 & & 64 & 64 & & 87 & 86  \\
 &  & E(n)  & 428 & 410 & & 399 & 371 & & 440 & 440 & & 393 & 395   \\ 
 & LD-OBF & Rej at interim & 16 & 24 & & 26 & 40 & & 11 & 11 & & 26 & 24  \\  
 &  & Rej at final & 73 & 77 & & 88 & 91 & & 66 & 66 & & 87 & 87   \\
 &  & Power & 78 & 82 & & 91 & 95 & & 70 & 70 & & 90 & 90 \\
 &  & E(n) & 466 & 450 & & 445 & 416 & & 478 & 478 & & 446 & 449 \\ 
\hline
\end{tabular}

\end{threeparttable}}
\end{table}

\section{Analysis of Neuroblastoma Study Data}
\label{sec5}

\begin{figure}
    \centerline{\includegraphics[scale=.35]{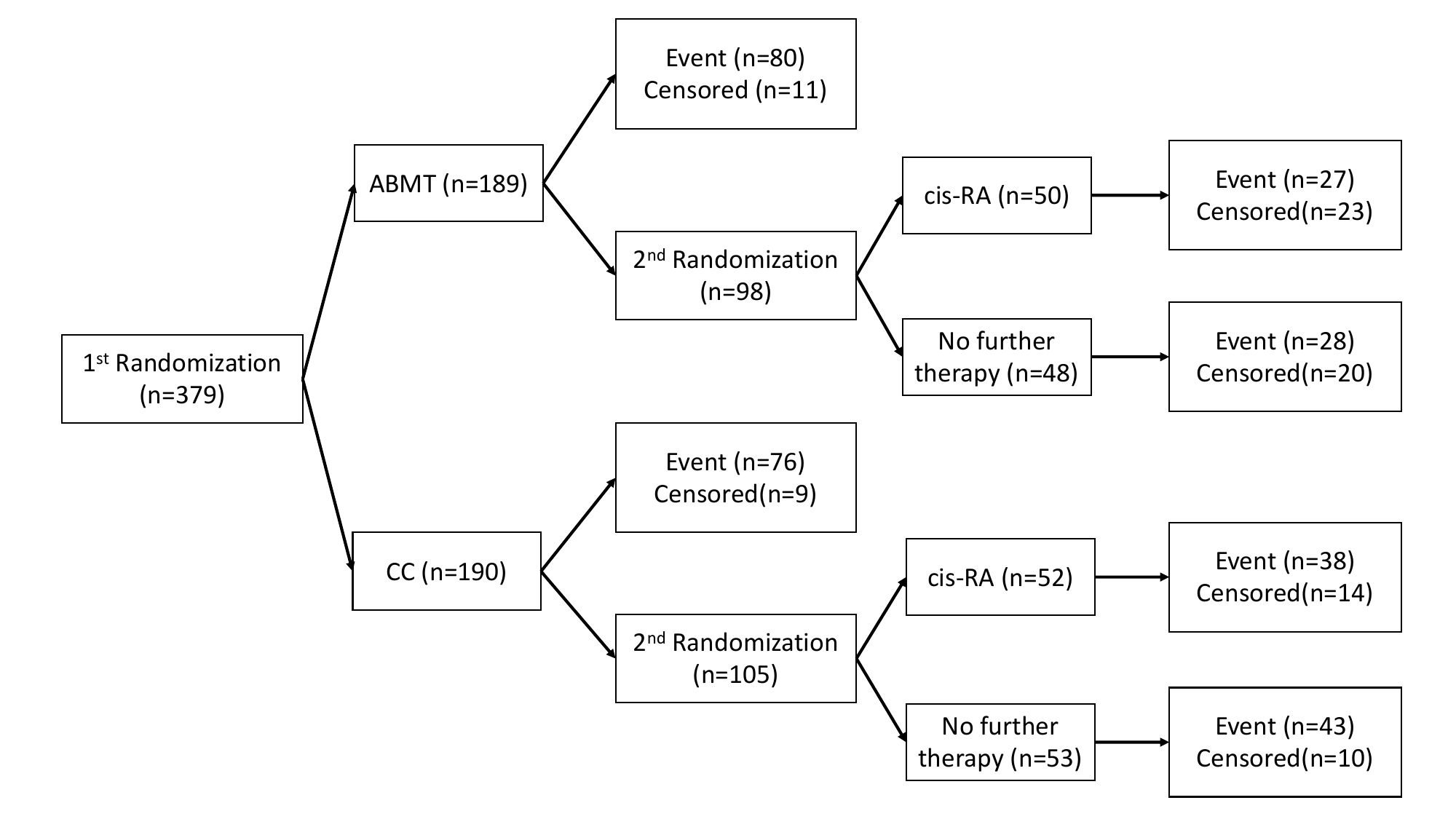}}
    \vspace{-.1in}
    \caption{Flowchart of Neuroblastoma Trial; Event, progression or death.  \label{SMART_Flowchart}}
\end{figure}

We applied the proposed interim monitoring method retrospectively to the Children's Cancer Group high-risk neuroblastoma study \citep{Matthay1999, Matthay2009}. In the study, 539 patients were recruited between Jan 1991 and Apr 1996. All patients were initially treated with chemotherapy, and 379 patients without disease progression after the initial treatment participated in the first-stage randomization. These patients were randomized to autologous bone marrow transplantation (ABMT), a combination of myeloablative chemotherapy, total-body irradiation,
and transplantation of autologous bone marrow
purged of cancer cells ($n=189$) or continuation chemotherapy ($n=190$). Among them, 203 patients without disease progression after the first-stage treatment were further randomized to 13-cis-retinoic acid (cis-RA) or no further therapy (102 in cis-RA and 101 no therapy). Non-responders to the first-stage treatment did not enter the second stage of the study, and hence this study is similar to the SMART2 design considered in the previous sections. The flowchart of the Neuroblastoma trial is shown in Figure \ref{SMART_Flowchart}. 

There are four embedded DTRs in this trial: 1) ABMT, cis-RA: treat with ABMT and then with cis-RA if no progression; 2) ABMT, none: treat with ABMT and receive no further therapy; 3) CC, cis-RA: treat with continuation chemotherapy and then with cis-RA if there was no disease progression; and 4) CC, none: treat with continuation chemotherapy and received no further therapy if no progression. 


We reanalyzed the Neuroblastoma trial to test if there was a significant difference in progression-free survival among the four DTRs by incorporating an interim analysis. Since the dataset did not contain enrollment time, to mimic the process of conducting an interim analysis planned a priori, we assumed the patients were enrolled uniformly in the pre-specified recruitment period (1991-1996), following the order of their study ID. Suppose the interim analysis is conducted after  50\% of the events are observed (1254 days, $n=$289). 

Under the null hypothesis, the Wald-type test statistic should follow a $\chi_3^2$ distribution.
The Pocock efficacy boundaries for this two-stage study, based on both the approximation and  error-spending approaches, are approximately 9.1 at both stages. In contrast, the OBF efficacy boundaries are approximately 11.5 and 8, respectively; details are in Web Table 5 of the supplementary material. At the interim analysis, the weighted log-rank and Tsiatis and Davidian's test statistics, 3.34 and 3.14, respectively, are below the critical values, suggesting  the trial should continue. The final test statistics are 5.80 (LR) and 6.56 (TD), which align with the interim decision not to reject the null hypothesis of no difference among the four DTRs.

\begin{figure}
    \centering
    \includegraphics[width=0.6\textwidth]{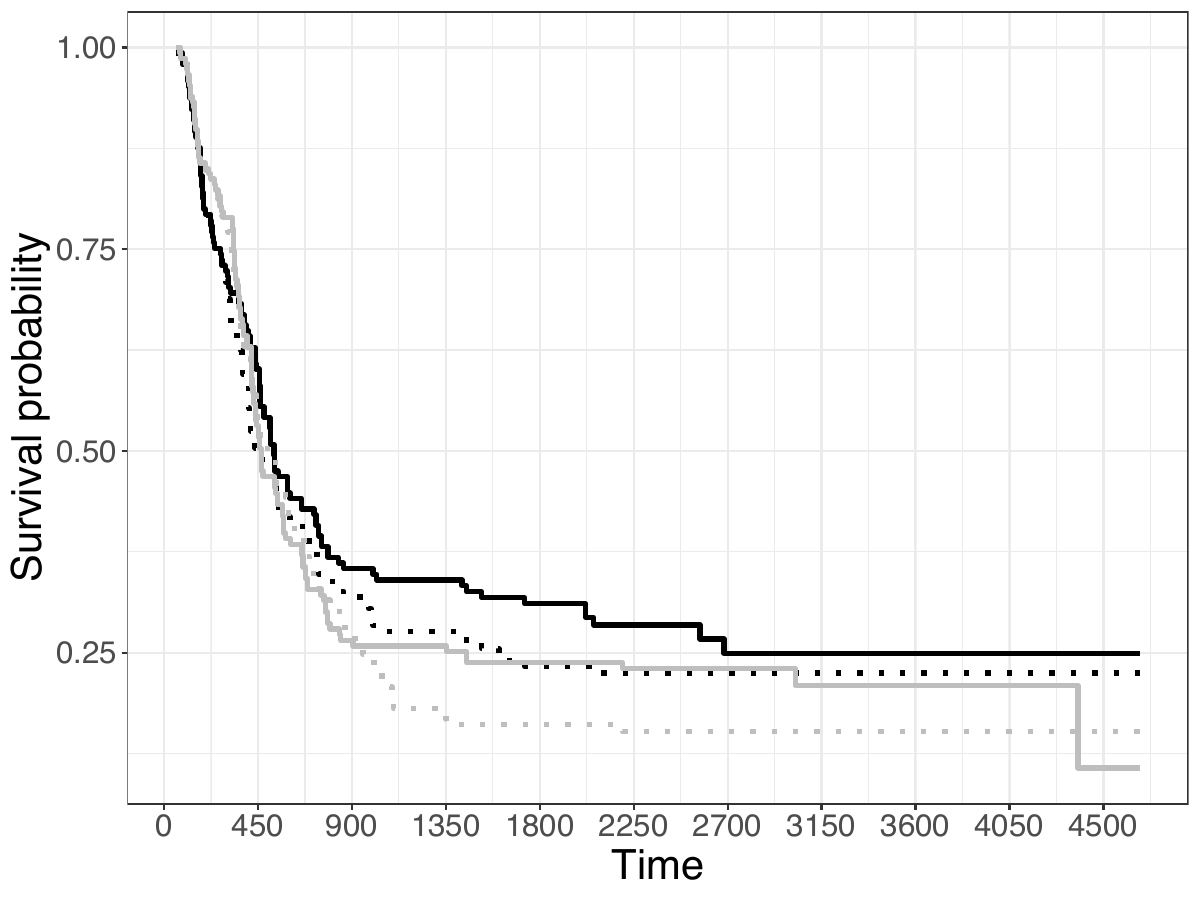}
\includegraphics[width=0.6\textwidth]{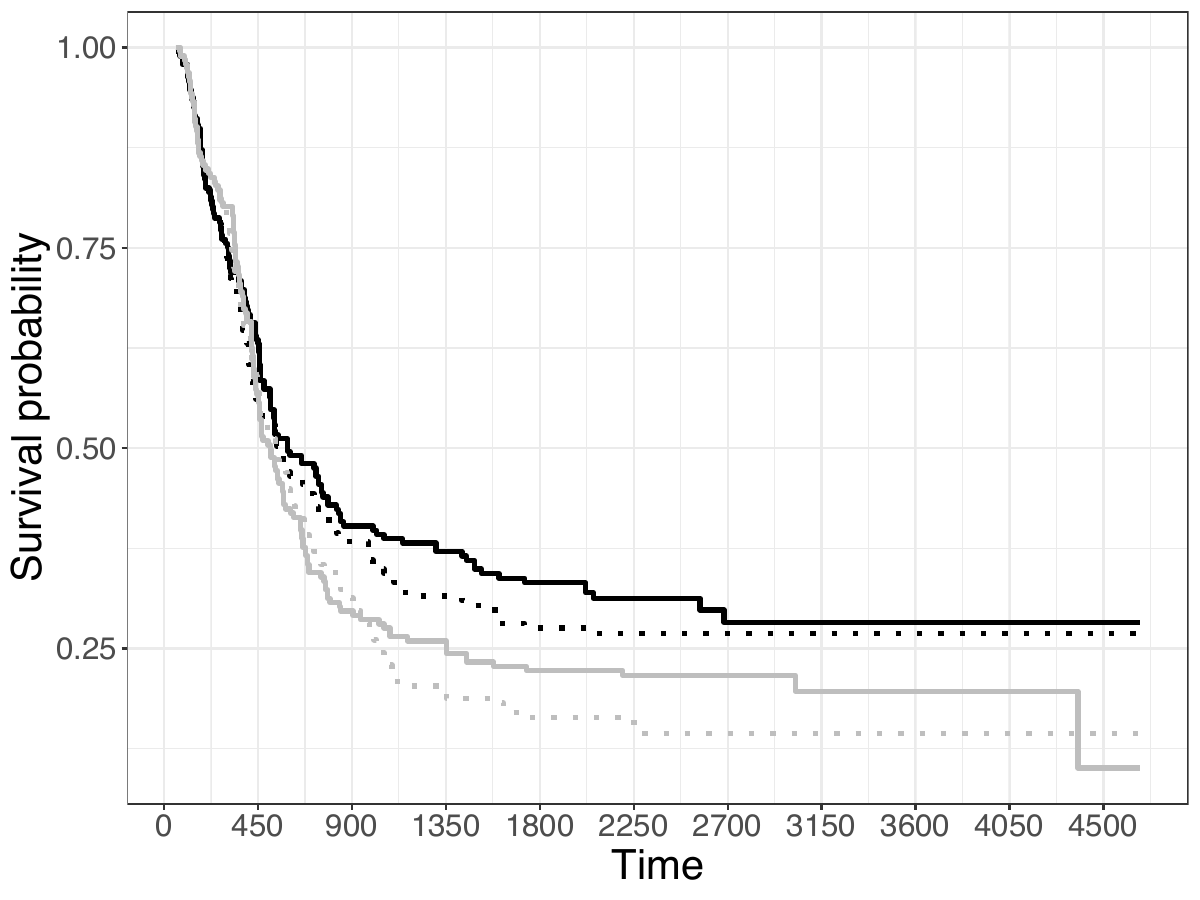}      
\caption{Survival Curves for 4 Treatment Strategies at interim (top panel) and final (bottom panel) analyses for the Neuroblastoma study. The black solid, black dotted, grey solid and grey dotted curves are survival curves for 4 treatment strategies: ABMT, cis-RA; ABMT, none; CC, cis-RA; CC, none.}
\label{Neuro_2Stage}
\end{figure}

We also estimated the survival curves for the four DTRs using the weighted risk set estimator of  \cite{Guo2005} at both the interim and final analyses, as shown in Figure \ref{Neuro_2Stage}. The median survival times for the four treatment strategies are also reported here. We observe differences in the four curves and median survival times (e.g., 526, 440, 447, and 497 days in the interim). However, these differences are not statistically significant due to small sample sizes and large variability. 
Reassuringly, weighted log-rank and Tsiatis and Davidian's tests lead to the same conclusion. 

\section{Discussion}
\label{sec6}

DTRs are widely used in managing chronic diseases, and SMART is instrumental in comparing embedded DTRs. In this paper, we proposed interim analysis procedures for SMARTs with survival outcomes based on weighted log-rank statistic and Tsiatis and Davidian's statistic. We also presented the statistical properties of the proposed test statistics and validated their use through rigorous simulation studies. Additionally, we applied the proposed methods on a neuroblastoma dataset to demonstrate the resource-saving advantage of the interim monitoring in SMARTs. Interim monitoring can potentially end the trial early without inflating type I error and maintaining statistical power.

One major issue with survival outcomes is that the covariance matrix of test statistics across analyses is not constant over time. To solve this issue, we adopted two approaches, the approximation approach and the error spending function approach. Both approaches worked well for various SMART designs in extensive simulations.

Two commonly used efficacy boundaries, the Pocock and OBF boundaries, and the Lan and Demets error spending methods were considered and evaluated in this work. The proposed interim analysis framework can easily incorporate other boundaries, such as those introduced by \cite{Haybittle1971}.
The method can also be extended to accommodate additional stages and a wider range of treatment options at each stage.


\backmatter


\section{Acknowledgements}

This work was partially funded through a Patient-Centered Outcomes Research Institute (PCORI) Award (ME-2021C3-24215). The statements in this work are solely the responsibility of the authors and do not necessarily represent the views of PCORI, its Board of Governors or the Methodology Committee. We thank all the stakeholder advisory board members of the study: Jordan Karp of the University of Arizona, Julie Bauman of the George washington University, Peter Thall of the MD Anderson Cancer Center, Douglus Landsitell of the University of Buffalo, and Gong Tang, Joyce Chang, and Meredith Wallace-Lotz of the University of Pittsburgh, for their input on the project.  This work was also partially supported by the University of Pittsburgh Center for Research Computing, RRID:SCR-022735, through the resources provided. Specifically, this work used the H2P cluster, which is supported by NSF award number OAC-2117681. \vspace*{-8pt}


\section{Supplementary Material}
The supplementary material contains available Web Appendices, Tables, and Figures referenced in Sections 4 and 5. \vspace*{-8pt}




\begin{thebibliography}{}

\bibitem[\protect\citeauthoryear{Bigirumurame, Uwimpuhwe, and
  Wason}{Bigirumurame et~al.}{2022}]{Bigirumurame2022}
Bigirumurame, T., Uwimpuhwe, G., and Wason, J. (2022).
\newblock Sequential multiple assignment randomized trial studies should report
  all key components: a systematic review.
\newblock {\em Journal of Clinical Epidemiology} {\bf 142,} 152--160.

\bibitem[\protect\citeauthoryear{Broglio, Stivers, and Berry}{Broglio
  et~al.}{2014}]{Broglio2014}
Broglio, K.~R., Stivers, D.~N., and Berry, D.~A. (2014).
\newblock Predicting clinical trial results based on announcements of interim
  analyses.
\newblock {\em Trials} {\bf 15,} 73--80.

\bibitem[\protect\citeauthoryear{Chakraborty and Murphy}{Chakraborty and
  Murphy}{2014}]{Chakraborty2014}
Chakraborty, B. and Murphy, S.~A. (2014).
\newblock Dynamic treatment regimes.
\newblock {\em Annu, Re. Stat. Appl.} {\bf 1,} 447--464.

\bibitem[\protect\citeauthoryear{Chen, Demets, and Lan}{Chen
  et~al.}{2003}]{Chen2003}
Chen, J., Demets, D., and Lan, G. (2003).
\newblock Monitoring mortality at interim analyses while testing a composite
  endpoint at the final analysis.
\newblock {\em Controlled Clinical Trials} {\bf 24,} 16--27.

\bibitem[\protect\citeauthoryear{Dickhaus and Royen}{Dickhaus and
  Royen}{2015}]{dickhaus2015}
Dickhaus, T. and Royen, T. (2015).
\newblock A survey on multivariate chi-square distributions and their
  applications in testing multiple hypotheses.
\newblock {\em Statistics} {\bf 49,} 427--454.

\bibitem[\protect\citeauthoryear{Ellenber, Fleming, and Demets}{Ellenber
  et~al.}{2002}]{Ellenberg2002}
Ellenber, S.~S., Fleming, T.~R., and Demets, D.~I. (2002).
\newblock {\em Data Monitoring Committees in Clinical trials}.
\newblock Willey, Chichester.

\bibitem[\protect\citeauthoryear{Freidlin, Korn, and Gray}{Freidlin
  et~al.}{2010}]{Freidlin2010}
Freidlin, B., Korn, E.~L., and Gray, R. (2010).
\newblock A general inefficacy interim monitoring rule for randomized clinical
  trials.
\newblock {\em Clinical Trials} {\bf 7,} 197--208.

\bibitem[\protect\citeauthoryear{Gu and Lai}{Gu and Lai}{1999}]{Gu1999}
Gu, M. and Lai, T.~L. (1999).
\newblock Determination of power and sample size in the design of clinical
  trials with failure-time endpoints and interim analyses.
\newblock {\em Controlled Clinical Trials} {\bf 20,} 423--438.

\bibitem[\protect\citeauthoryear{Guo and Tsiatis}{Guo and
  Tsiatis}{2005}]{Guo2005}
Guo, X. and Tsiatis, A.~A. (2005).
\newblock weighted risk set estimator for survival distributions in two-stage
  randomization designs with censored survival data.
\newblock {\em The International Journal of Biostatistics} {\bf 1,} 1--15.

\bibitem[\protect\citeauthoryear{Harrington and Fleming}{Harrington and
  Fleming}{1982}]{Harrington1982}
Harrington, D. and Fleming, T. (1982).
\newblock A class of rank test procedures for censored survival data.
\newblock {\em Controlled Clinical Trials} {\bf 69,} 553--566.

\bibitem[\protect\citeauthoryear{Haybittle}{Haybittle}{1971}]{Haybittle1971}
Haybittle, J.~L. (1971).
\newblock Repeated assessment of results in clinical trials of cancer
  treatment.
\newblock {\em The British Journal of Radiology} {\bf 44,} 793--797.

\bibitem[\protect\citeauthoryear{Jennsion and Turnbull}{Jennsion and
  Turnbull}{1990}]{Jennsion1990}
Jennsion, C. and Turnbull, B. (1990).
\newblock Statistical approaches to interim monitoring of medical trials: A
  review and commentary.
\newblock {\em Statistical Science} {\bf 5,} 299--317.

\bibitem[\protect\citeauthoryear{Kim and Tsiatis}{Kim and
  Tsiatis}{1990}]{Kim1990}
Kim, K. and Tsiatis, A. (1990).
\newblock Study duration for clinical trials with survival response and early
  stopping rule.
\newblock {\em Biometrics} {\bf 46,} 81--92.

\bibitem[\protect\citeauthoryear{Lan and Demets}{Lan and
  Demets}{1983}]{Lan1983}
Lan, K. and Demets, D.~L. (1983).
\newblock Discrete sequential boundaries for clinical trials.
\newblock {\em Biometrics} {\bf 70,} 659--663.

\bibitem[\protect\citeauthoryear{Lavori}{Lavori}{2008}]{Lavori2008}
Lavori, P.W.~andDawson, R. (2008).
\newblock Adaptive treatment strategies in chronic disease.
\newblock {\em Annual Review of Medicine} {\bf 59,} 443--453.

\bibitem[\protect\citeauthoryear{Li, Valenstein, Pfeiffer, and Ganoczy}{Li
  et~al.}{2014}]{Li2014}
Li, Z., Valenstein, M., Pfeiffer, P., and Ganoczy, D. (2014).
\newblock A global logrank test for adpative treatment strategies based on
  observational studies.
\newblock {\em Statistics in Medicine} {\bf 33,} 760--771.

\bibitem[\protect\citeauthoryear{Manschot, Laber, and Davidian}{Manschot
  et~al.}{2023}]{Manschot2023}
Manschot, C., Laber, E., and Davidian, M. (2023).
\newblock Interim monitoring of sequential multiple assignment randomized
  trials using partial information.
\newblock {\em Biometrics} {\bf 79,} 2881--2894.

\bibitem[\protect\citeauthoryear{Matthay, Reynolds, Seeger, Shimada, Adkins,
  Haas-Kogan, Gerbing, London, and Villablanca}{Matthay
  et~al.}{2009}]{Matthay2009}
Matthay, K.~K., Reynolds, C.~P., Seeger, R.~C., Shimada, H., Adkins, E.~S.,
  Haas-Kogan, D., Gerbing, R.~B., London, W.~B., and Villablanca, J.~G. (2009).
\newblock Long-term results for children with high-risk neuroblastoma treated
  on a randomized trial of myeloablative therapy followed by 13-cis-retinoic
  acid: A children's oncology group study.
\newblock {\em Journal of Clinical Oncology} {\bf 27,} 1007--1013.

\bibitem[\protect\citeauthoryear{Matthay, Villablanca, Seeger, Stram, Harris,
  Ramsay, Swift, Shimada, Black, Brodeur, Gerbing, and Reynolds}{Matthay
  et~al.}{1999}]{Matthay1999}
Matthay, K.~K., Villablanca, J.~G., Seeger, R.~C., Stram, D.~O., Harris, R.~E.,
  Ramsay, N.~K., Swift, P., Shimada, H., Black, C.~T., Brodeur, G.~M., Gerbing,
  R.~B., and Reynolds, C.~P. (1999).
\newblock Treatment of high-risk neuroblastoma with intensive chemotherapy,
  radiotherapy, autologous bone marrow transplantation, and 13-cis-retinoic
  acid.
\newblock {\em The New England Journal of Medicine} {\bf 341,} 1165--1173.

\bibitem[\protect\citeauthoryear{Murphy}{Murphy}{2005}]{Murphy2005}
Murphy, S.~A. (2005).
\newblock An experimental design for the development of adaptive treatment
  strategies.
\newblock {\em Statistics in Medicine} {\bf 24,} 1455--1481.

\bibitem[\protect\citeauthoryear{Nahum-Shani, Ertefaie, Lu, Lyncg, Mckay,
  Oslin, and Almirall}{Nahum-Shani et~al.}{2017}]{Nahum-Shani2017}
Nahum-Shani, I., Ertefaie, A., Lu, X., Lyncg, K.~G., Mckay, J., Oslin, D., and
  Almirall, D. (2017).
\newblock A smart data analysis method for constructing adaptive treatment
  strategies for substance use disorders.
\newblock {\em Addiction} {\bf 112,} 901--909.

\bibitem[\protect\citeauthoryear{O'Brien and Fleming}{O'Brien and
  Fleming}{1979}]{Brien1979}
O'Brien, P.~C. and Fleming, T.~R. (1979).
\newblock A multiple testing procedure for clinical trials.
\newblock {\em Biometrics} {\bf 35,} 549--556.

\bibitem[\protect\citeauthoryear{Pocock}{Pocock}{1977}]{Pocock1977}
Pocock, S.~J. (1977).
\newblock Group sequential methods in the design and analysis of clinical
  trials.
\newblock {\em Biometrika} {\bf 64,} 191--199.

\bibitem[\protect\citeauthoryear{Shen and Cai}{Shen and Cai}{2003}]{Shen2003}
Shen, Y. and Cai, J. (2003).
\newblock Sample size reestimation for clinical trials with censored survival
  data.
\newblock {\em Journal of the American Statistical Association} {\bf 98,}
  418--426.

\bibitem[\protect\citeauthoryear{Tsiatis}{Tsiatis}{1982}]{Tsiatis1982}
Tsiatis, A. (1982).
\newblock Repeated significance testing for a general class of statistics used
  in censored survival analysis.
\newblock {\em Journal of the American Statistical Association} {\bf 77,}
  855--861.

\bibitem[\protect\citeauthoryear{Tsiatis and Davidian}{Tsiatis and
  Davidian}{2024}]{Tsiatis2024}
Tsiatis, A.~A. and Davidian, M. (2024).
\newblock A generalized logrank-type test for comparison of treatment regimes
  in sequential multiple assignment randomized trials.
\newblock {\em Biometrics. arXiv preprint arXiv:2403.16813} .

\bibitem[\protect\citeauthoryear{Wu, Wang, and Wahed}{Wu et~al.}{2021}]{wu2021}
Wu, L., Wang, J., and Wahed, A. (2021).
\newblock Interim monitoring in sequential multiple assignment randomized
  trials.
\newblock {\em Biometrics} pages 1--13.

\end{thebibliography}
\providecommand{\newblock}{}

\label{lastpage}

\end{document}